 \documentclass[smallabstract,smallcaptions]{dccpaper}

\usepackage{epsfig}
\usepackage{citesort}
\usepackage{amsmath}
\usepackage{amssymb}
\usepackage{color}
\usepackage{url}
\usepackage{subcaption}
\usepackage{algorithm}
\usepackage{algorithmic}

\newlength{\figurewidth}
\newlength{\smallfigurewidth}

\makeatletter
\newcommand*{\rom}[1]{\expandafter\@slowromancap\romannumeral #1@}
\makeatother
\newcommand\given[1][]{\:#1\vert\:}

\begin{document}

\title
{\large
\textbf{A Bayesian Approach to Block Structure Inference in AV1-based Multi-rate Video Encoding}
}

\author{%
Bichuan Guo$^{\ast}$, Xinyao Chen$^{\ast}$, Jiawen Gu$^{\ast}$, Yuxing Han$^{\dag}$, Jiangtao Wen$^{\ast}$\\[0.5em]
{\small\begin{minipage}{\linewidth}\begin{center}
\begin{tabular}{ccc}
$^{\ast}$Tsinghua Univeristy & \hspace*{0.5in} & $^{\dag}$South China Agricultural Univeristy \\
Beijing, 100084, China && Guangzhou, Guangdong, 510642, China\\
\url{jtwen@tsinghua.edu.cn} && \url{yuxinghan@scau.edu.cn}
\end{tabular}
\end{center}\end{minipage}}
}

\maketitle
\thispagestyle{empty}

\begin{abstract}
Due to differences in frame structure, existing multi-rate video encoding algorithms
cannot be directly adapted to encoders utilizing special reference frames such as AV1
without introducing substantial rate-distortion loss.
To tackle this problem, 
we propose a novel bayesian block structure inference model inspired by a modification to an HEVC-based algorithm. 
It estimates the posterior probabilistic distributions of block partitioning, 
and adapts early terminations in the RDO procedure accordingly.
Experimental results show that the proposed method provides flexibility for controlling  
the tradeoff between speed and coding efficiency, and can achieve an average time saving of $36.1\%$ (up to $50.6\%$) with negligible 
bitrate cost.
\end{abstract}

\Section{Introduction}

Video content accounts for the majority of all internet traffic, 
and has been  growing steadily \cite{Cisco17}.
Due to the dynamic nature of public network conditions,
adaptive streaming is widely used by video content providers \cite{Oyman12},
where the same video is encoded to different rates (and correspondingly qualities, denote as a variable $Q$), 
with a client-side algorithm adaptively requests a version based on network conditions.

On the other hand, AOMedia Video 1 (AV1 for short) is an open, 
royalty-free video coding format developed by the Alliance for Open Media \cite{Wiki-av1}\cite{Git-av1}.
Based on its predecessor VP9, 
AV1 utilizes numerous new coding tools to achieve 
cutting-edge coding efficiency.
As the coding complexity increases, 
real time encoding becomes a challenge for adaptive live streaming, and it is thus of great 
interest to reduce the overall complexity without substantial degradation to the coding efficiency.

In addition to the challenges inherent in real time encoding a single AV1 stream, encoding
an input video to multiple rates in parallel is another challenge of great interests to adaptive
video streaming. Given that the rate-distortion optimizations (RDO) in multiple encodings of  
the same video is correlated, parameters and intermediate results from the different  RDO procedures can be
shared among the processes. Such parameters and intermediate results include
prediction modes, motion vectors/intra modes, and block structures \cite{Schroeder17}. 
It has been shown \cite{Schroeder17}\cite{Praeter15}\cite{Schroeder15} that the encoder complexity 
can be significantly reduced by considering block structures alone.

In this paper, we propose a bayesian inference model for block structures in video codecs using special long-term reference frames.
It exploits the statistical correlation between encoding processes for the same video input with different target quality levels 
and identical input/output spatial resolutions.
The AV1 codec
is used as a benchmark tool to demonstrate the effectiveness of our model without loss of generality, 
since the proposed model can be applied to other codecs of similar principles.

The rest of the paper is organized as follows.
Related work is presented in Section \rom{2}.
A study on the statistical behavior of block structure decisions of the AV1 codec is given in Section \rom{3}.
Adaptation of an existing algorithm from HEVC to AV1 and its improvement are discussed in Section \rom{4},
based on which, a detailed description and analysis of 
our proposed bayesian inference model are given in Section \rom{5},
followed by experimental results in Section \rom{6}.
Finally, Section \rom{7} concludes the paper.

\Section{Related Work}

Many efforts have been dedicated to reduce the complexity of video encoding.
\cite{Zhao06}\cite{Zhao06-2} proposed a wavefront parallel processing method that exploits spatial independencies in H.264/AVC,
which have since been adopted by the x264 open source H.264 encoder \cite{x264}.
\cite{Wen16} improves it by jointly considering temporal independencies.
These methods accelerate video encoding on multi-core systems without reducing its complexity.
Another approach \cite{Shen13}\cite{Chen15} is to use transcoding \cite{Ahmad05} for multiple encoding of
the same input. Because the RDO mode decisions and motion estimation processes are bitrate dependent,
and different video coding formats such as H.264 and H.265 are very different,  transcoding by simply re-quantizating 
the residual calculated using the mode and motion information obtained from the RDO
for a encoding format and a different rate may introduce significant quality losses, both from not
fully utilizing encoding tools in a target format, and from sub-optimal RDO decisions.

Simultaneous encoding of the same input,
also called multi-rate encoding \cite{Li03}, 
reduces the overall complexity while retaining high fidelity.
The encoder performs full encoding on a chosen instance as a \textit{reference instance},
while other instances (referred to as \textit{local instances})
consult the reference instance to infer their optimal RDO decisions.
\cite{Finstad11} proposed a preliminary framework which merely copies decisions from the reference instance,
resulting in considerable rate-distortion (RD) loss.
\cite{Schroeder17}\cite{Schroeder15} refined this framework by
searching in a pruned RDO recursion tree,
reducing the RD loss to a negligible level. 
However, as shown later in this paper,
this method cannot be directly migrated to AV1 without introducing substantial performance degradation.
\cite{Praeter15} proposed an ensemble learning method to predict optimal block structures in HEVC multi-rate encoding.

Bayesian inference is widely used in video encoding. \cite{Hu15}\cite{Shen12} used bayesian models to accelerate 
skip mode and CU size decisions in HEVC. 
\cite{Cai16} uses a bayesian model to accelerate prediction mode decisions in x265 multi-rate encoding.
There has been very little reported work that has considered block structure decisions with RDO 
of both the reference instance and the local instance in multi-rate encoding.

As its main contribution, 
the current paper proposes a joint probabilistic model to infer optimal block structure decisions.
For a particular local instance, 
its past block structure decisions will be recorded to update the inference model, 
which in turn computes the probabilistic distribution of its optimal block structure decisions with the aid of the reference instance.

\Section{Block Structures in AV1}


We introduce some terms that will be used throughout this paper. 
The \textit{depth} of a $w \times h$ block is defined to be
$\min (\log_2 64/w, \log_2 64/h)$,
i.e. the difference between the base $2$ logarithms of 64 and the length of the longer edge.
A block is \textit{prime} if it is not split in the optimal block structure.
Two blocks \textit{overlap} if they have non-empty intersection, 
regardless of their parent frames.
A block $B$ has a \textit{split degree} of $d_f$ \textit{in frame} $f$,
if $d_f$ is the maximum depth of the prime blocks in frame $f$ that overlap with $B$.
The \textit{remote frame} of a block $B$ refers to the frame in the reference instance that is identical to the parent frame of $B$.

In AV1, the block structure optimization starts from 64$\times$64 blocks.
A 2N$\times$2N block can be partitioned into four N$\times$N blocks (referred to as \textit{4-split}), 
two N$\times$2N blocks, two 2N$\times$N blocks, or no partition at all.
A non-square block does not split.
Each frame is associated with a \textit{q-index}
that indicates the position of the frame's base quantization parameter (QP) in the lookup table.
Periodically some \textit{special frames} are selected for long-term reference and are given relatively 
lower QP's, namely intra-frames, golden-frames, and altref-frames, to improve quality.

\SubSection{Common Test Settings}

\begin{table}[tp]
\begin{center}
\caption{\label{tab:av1-configuration}%
Common configurations of the AV1 encoder}
{
\renewcommand{\baselinestretch}{1}\footnotesize
\begin{tabular}{|c|c|c|c|c|c|}
\hline
Parameter &Value &Parameter &Value &Parameter &Value \\
\hline
cpu-used &0 &kf-min-dist &0 &bit-depth &8 \\
end-usage &q &kf-max-dist &9999  &auto-altref &1 \\
pass &1 &kf-mode &1 &drop-frame &0 \\
\hline
\end{tabular}}
\end{center}
\end{table}

In this paper we use the AV1 codec v0.1.0 \cite{Git-av1-v0.1.0} running on Ubuntu 15.04 with  a 2.40 GHz Intel Xeon E5-2695v2 CPU and 64GB RAM.
The common configuration that will be used throughout this paper is given in Table~\ref{tab:av1-configuration}.
A set of eight test sequences with different spatial resolutions was selected 
(see Table~\ref{tab:simple-rdo-pruning}).
RD performance and encoding time is measured by BD-rate \cite{BD01}\cite{BD08}
and CPU time. 

\SubSection{Statistical Behavior}

\begin{figure}[t]
\begin{center}
\epsfig{width=2.5in,file=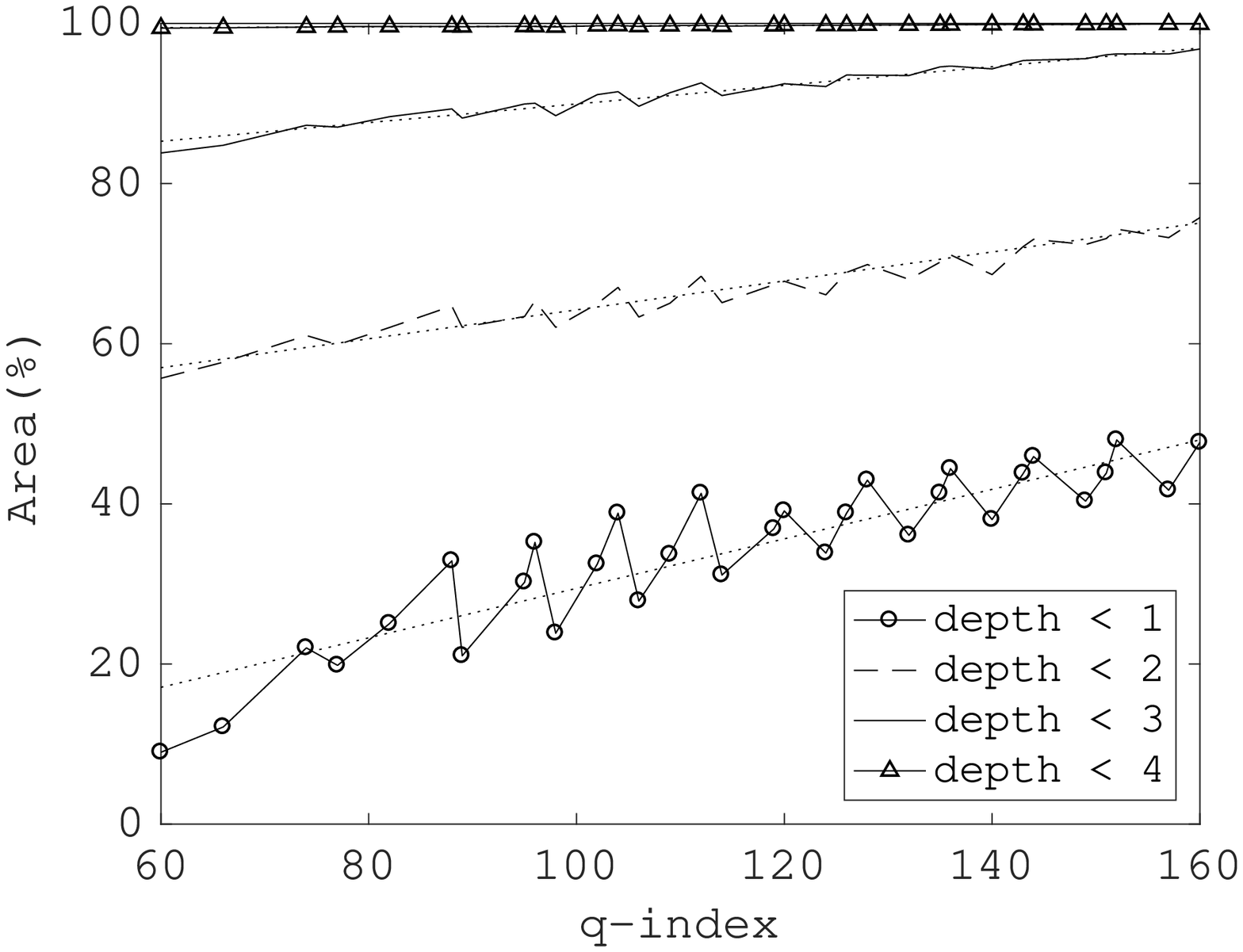} \\
{\small (a)} \\
\begin{tabular}{cc}
\epsfig{width=2.5in,file=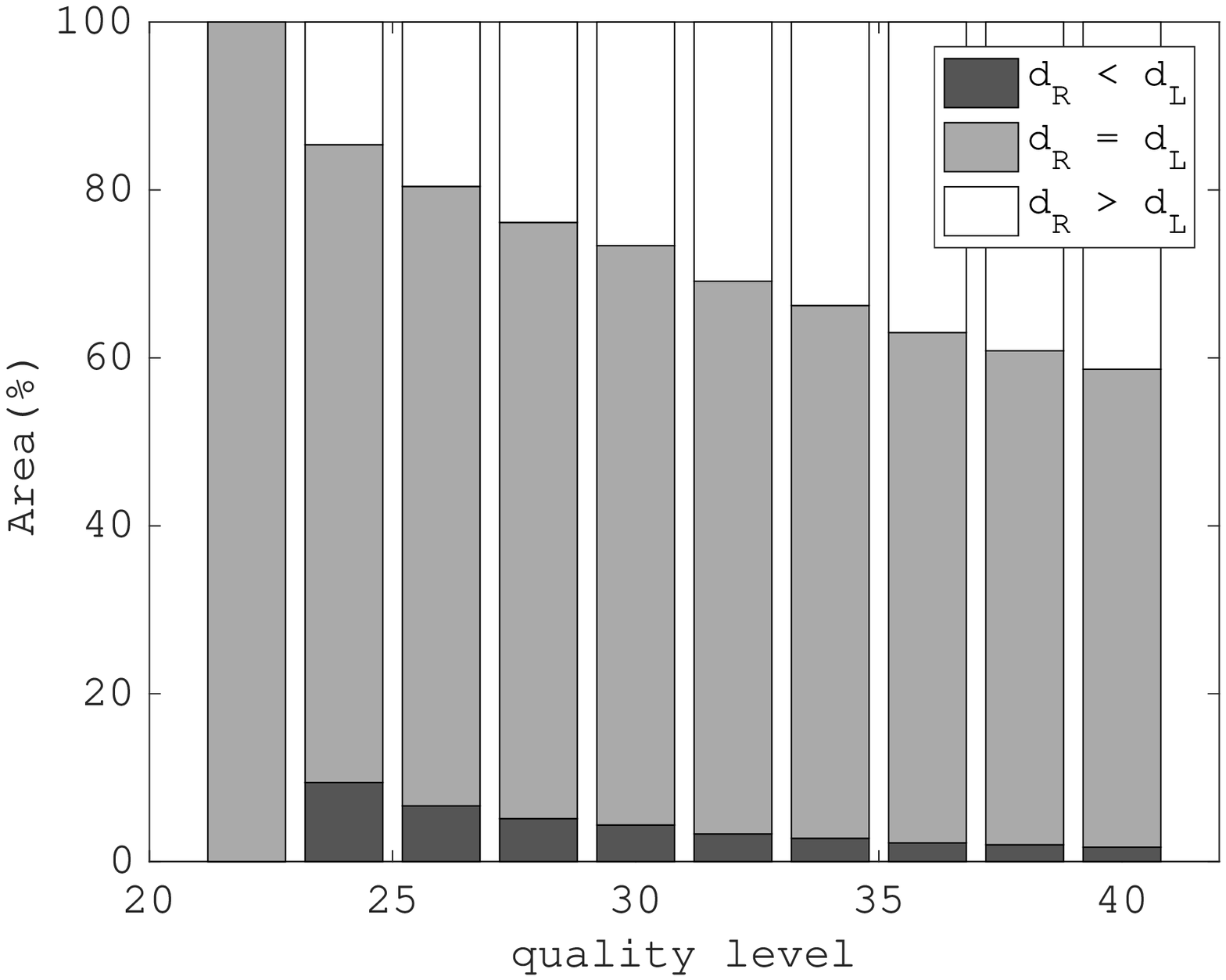} &
\epsfig{width=2.3in,file=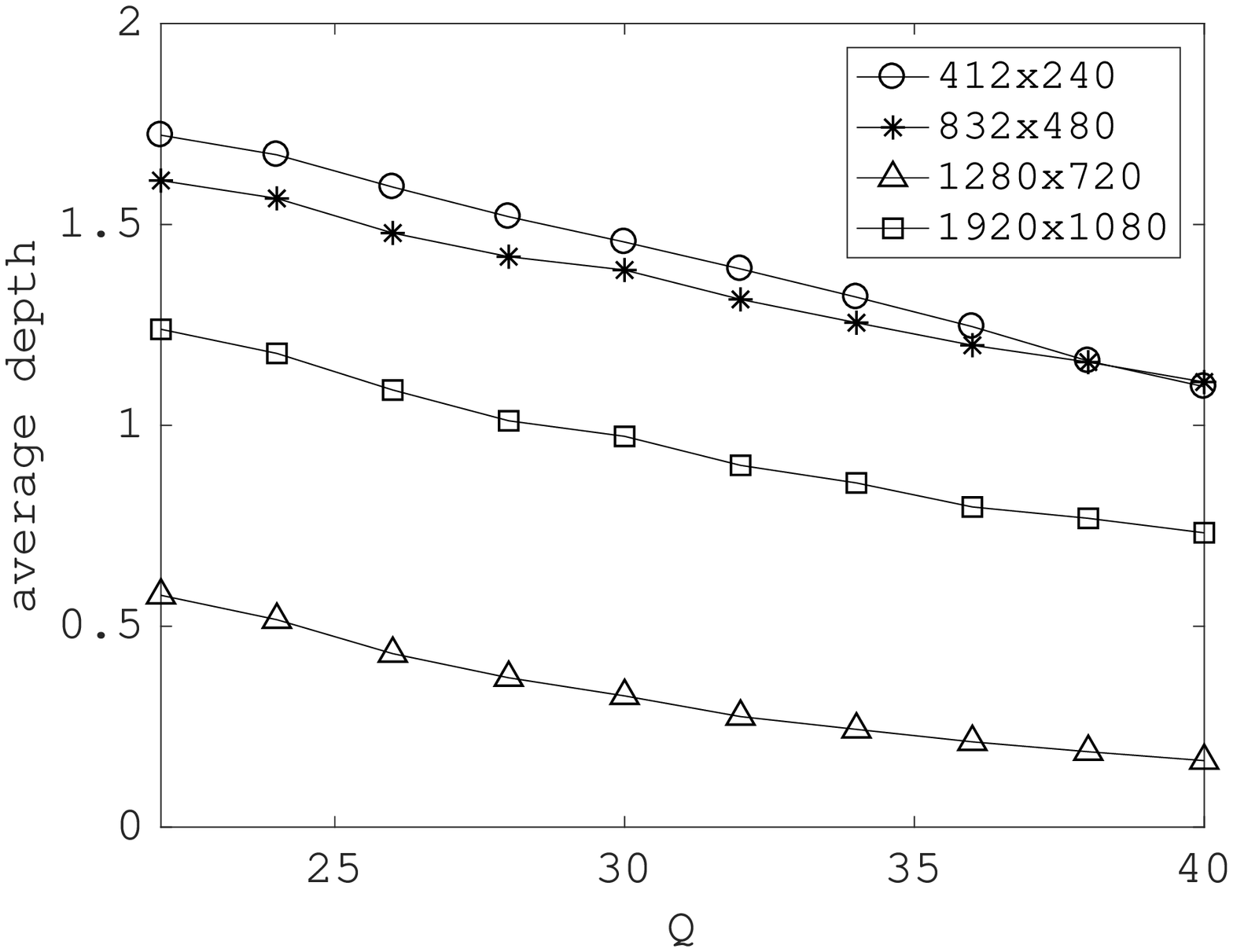} \\
{\small (b)} & {\small (c)}
\end{tabular}
\end{center}
\caption{\label{fig:block-structure-statistics}%
Statistical behavior of optimal block structures}
\end{figure}

Fig.~\ref{fig:block-structure-statistics}(a) shows the percentage of total area for prime blocks 
as a function of their max depths, averaged across all non-special frames with identical q-indices.
Note that we use q-indices rather than the actual QP's for better regression.
50 frames of the sequence \textit{BasketballDrill} were tested with quality levels set to even values from 22 to 40.
Observe that prime blocks in frames with larger q-indices tend to have smaller depths, 
in other words, partitions occur less often, as expected. 

The linear approximations, shown as dotted lines in the figure,
give the prior probability of 4-splitting a square block. 
Denote the linear approximations as $a_d(q), d=1,...,4$ that 
map q-indices to area percentages. 
For a square block $B$ in frame $f$ whose depth is $d$, 
let $d_{f'}$ be the split degree of $B$ in an arbitrary frame $f'$ with q-index $q$.
We roughly estimate $P(d_{f'} < d)$ with $a_d(q)$, by the definition of $d_{f'}$, we have
\begin{align} \label{eqn:prior}
P(d_{f'} > d \given d_{f'} > d - 1) = \begin{cases}\displaystyle\frac{1 - a_{d+1}(q)}{1 - a_d(q)} & \text{if } d > 0, \\
1 - a_{d+1}(q) & \text{if } d = 0. \end{cases}
\end{align} 
Now let $f' = f$. We see that $d_{f'} = d_f \ge d$,  
thus (\ref{eqn:prior}) is exactly the probability of $B$ 4-splitting.
We will see later that this prior, although established using a single sequence,
is robust and applicable to various different sequences for the purpose of this paper.

We now turn to the correlation of optimal block structures between the reference
and local instances. Using the same test settings in Fig.~\ref{fig:block-structure-statistics}(a), 
the instance with the best $Q=22$ is selected as the reference instance. 
Let $d_L$ be the depth of an arbitrary prime block $B$ in frame $f_L$ of a local instance,
$f_R$ be the remote frame of $B$,
and $d_R$ be the split degree of $B$ in frame $f_R$.
Fig.~\ref{fig:block-structure-statistics}(b) shows the percentage in total area  of prime blocks
that satisfy $d_R < d_L$, $d_R = d_L$ or $d_R > d_L$, 
averaged across all encoder instances with identical Q's.
As expected, when $Q=22$ all blocks have $d_R = d_L$,
since the local instance is identical to the reference instance.
The proportion of the case $d_R < d_L$ is always below $10\%$, 
and is even lower as $Q$ increases,
while other cases are of notable proportions.
This confirms a result for HEVC as reported in \cite{Schroeder15}.
For future reference, the average depths of prime blocks for each Q and each spatial resolution are 
also recorded and shown in Fig.~\ref{fig:block-structure-statistics}(c).

\Section{Improved RDO Pruning}

We showed in the previous section that
by choosing the instance with the best quality as the reference instance, the depth of
a prime block rarely exceeds its split degree in its remote frame. 
\cite{Schroeder15} proposed a multi-rate encoding algorithm based on a similar observation in HEVC.
Specifically, 
when the depth of a block has reached its split degree in its remote frame,
the local instance executes early termination, i.e. 4-split is not searched in RDO for the current block,
as it is unlikely to be the optimal choice.
We implemented this algorithm in the AV1 codec, 
whose results are shown in Table~\ref{tab:simple-rdo-pruning} as \textit{original}.
200 frames of each test sequence were run with $Q=22,27,32,37,42$,
choosing $Q=22$ as the reference instance.
The time saving is measured by the average time saving of all local instances,
with the time cost of the reference instance excluded in the calculations so as
to avoid strong dependency on the number of local instances. This is reasonable as
the time cost of a single reference instance will become insignificant when the number of local instances 
is sufficiently large.

\cite{Schroeder15} reported a time saving of 46\%, 
and a 0.62\% BD-rate on \textit{BasketballPass}, a significant better tradeoff 
than our results for AV1.  We believe that the difference resulted from the structural difference between 
AV1 and HEVC on the frame level.
The special frames in AV1 play a major role in improving its coding efficiency, as shown in our next experiment. 
When skipping part of the RDO recursion tree and opting for suboptimal decisions,
the overall coding efficiency severely deteriorated. To mitigate this problem,
we propose to improve the algorithm by fully encoding special frames
in local instances. 
Using the same settings specified above, 
the results are shown in Table~\ref{tab:simple-rdo-pruning} as \textit{improved}.
As a trade-off between speed and quality,
the coding efficiency in terms of BD-rate is much more plausible.

\begin{table}[tp]
\begin{center}
\caption{\label{tab:simple-rdo-pruning}%
Experimental results of the RDO pruning algorithms}
{
\renewcommand{\baselinestretch}{1}\footnotesize
    \centering
        \begin{tabular}{|l|c|c|c|c|}
        \hline
         & \multicolumn{2}{|c|}{BD-rate} &\multicolumn{2}{|c|}{$\Delta T$} \\
        \cline{2-5}
        Sequence &original &improved &original &improved  \\
        \hline
        BasketballPass (412$\times$240)  & 2.47\% & 0.15\% & -31.3\% & -18.9\% \\
        BlowingBubbles (412$\times$240)  & 3.80\% & 0.25\% & -32.3\% & -21.1\% \\
        BQSquare (412$\times$240)        & 2.04\% & 0.05\% & -27.2\% & -17.6\% \\
        \hline
        BasketballDrill (832$\times$480) & 3.75\% & 0.16\% & -47.8\% & -30.6\% \\
        PartyScene (832$\times$480)      & 2.32\% & 0.08\% & -30.5\% & -19.8\% \\
        \hline
        FourPeople (1280$\times$720)     & 4.09\% & 0.10\% & -58.6\% & -40.2\% \\
        Johnny (1280$\times$720)         & 4.13\% & 0.09\% & -60.9\% & -42.4\% \\
        \hline
        Kimono (1920$\times$1080)        & 1.61\% & 0.24\% & -50.9\% & -37.5\% \\
        \hline
        \textbf{Average}                & \textbf{3.03\%} & \textbf{0.14\%} & \textbf{-42.4\%} & \textbf{-28.5\%} \\
        \hline
        \end{tabular}
}
\end{center}
\end{table} 

\Section{A Bayesian Inference Model}
The last experiment has demonstrated the necessity for keeping 
block structures of special frames optimal when conducting expedited AV1 encoding.
As we run exhaustive search on special frames, 
their optimal block structures become reliable evidence for inference,
which will be thoroughly discussed in this section.

Let us assume that $f$ is a non-special frame in a local instance $L$, 
$B$ is a block in $f$ that has a depth $d$ and can be 4-split,
$f_L$ is the latest special frame in $L$ preceding $f$,
and $f_R$ is the remote frame of $B$.
Let $d_{f'}$ denote the split degree of $B$ in any arbitrary frame $f'$.
Since $f_L$ and $f_R$ are both fully encoded,
$d_{f_L}$ and $d_{f_R}$ are available.

The key in our multi-rate encoding algorithm is estimating the probability that
$B$ is 4-split in the optimal block structure, knowing $d_{f_L}$ and $d_{f_R}$,
i.e. $P(d_f > d \given d_{f_L}, d_{f_R})$,
as $f$, $f_L$ and $f_R$ exhibit strong similarities.
Since (\ref{eqn:prior}) gives the prior probability $p_0 = P(d_f > d)$, 
by way of Bayes' theorem
\begin{align} \label{eqn:posterior}
P(d_f > d \given d_{f_L}, d_{f_R}) = \frac{P(d_{f_L}, d_{f_R} \given d_f > d) p_0}
{P(d_{f_L}, d_{f_R} \given d_f > d) p_0 + P(d_{f_L}, d_{f_R} \given d_f = d) (1-p_0)}.
\end{align}
Note that $d_f \ge d$ is always true. 
This bayesian model allows us to incorporate our prior knowledge about $p_0$ into inference.
For each depth $d$, we create two 2D tables, namely $T_d^+$ and $T_d^-$,
which record the number of occurrences for the pair $(d_{f_L}, d_{f_R})$
when $d_f > d$ and $d_f = d$, respectively.
$P(d_{f_L}, d_{f_R} \given d_f)$ is therefore approximated by
\begin{align} \label{eqn:likelihood}
P(d_{f_L}, d_{f_R} \given d_f > d) = \frac{T_d^+(d_{f_L},d_{f_R})}{\displaystyle\sum_{ i, j }T_d^+(i, j)},~
P(d_{f_L}, d_{f_R} \given d_f = d) = \frac{T_d^-(d_{f_L},d_{f_R})}{\displaystyle\sum_{ i, j }T_d^-(i, j)}.
\end{align}
We substitute (\ref{eqn:likelihood}) into (\ref{eqn:posterior}) to obtain the posterior probability $p$. 
Given $\tau_1, \tau_2 \in (0,1)$ and a uniform random variable $X\sim \mathcal{U}(0,1)$, 
the RDO procedure with regard to $B$ is described by the pseudocode.
Intuitively, $\tau_1$ is a threshold determining if $p$ is sufficiently small,
and $\tau_2$ is the sampling frequency for running exhaustive search even if $p$ is small.
Early termination is not used if $p$ is not sufficiently small. 
Even when it is used, we shall still occasionally run full encoding to keep the likelihoods unbiased.

\begin{algorithm}[tp]
\renewcommand{\baselinestretch}{1}\footnotesize
\caption{Proposed RDO procedure}
\begin{algorithmic}[1]
\REQUIRE $B$, $\tau_1$, $\tau_2$, $X$, $T_d^+$, $T_d^-$, $d_{f_L}$, $d_{f_R}$
\STATE compute $p=P(d_f>d\given d_{f_L}, d_{f_R})$ by (\ref{eqn:posterior})(\ref{eqn:likelihood})
\IF{$p \le \tau_1$ \AND $X \ge \tau_2$}
\STATE do not attempt to 4-split $B$ (early termination)
\ELSE
\IF{$p > \tau_1$}
\STATE $k \leftarrow 1$
\ELSE [$p \le \tau_1, X < \tau_2$]
\STATE $k \leftarrow \tau_2^{-1}$
\ENDIF
\STATE run ordinary RDO to determine if 4-splitting $B$ is optimal
\IF{4-splitting $B$ is optimal}
\STATE $T_d^+(d_{f_L},d_{f_R}) \leftarrow T_d^+(d_{f_L},d_{f_R}) + k$  
\ELSE
\STATE $T_d^-(d_{f_L},d_{f_R}) \leftarrow T_d^-(d_{f_L},d_{f_R}) + k$ 
\ENDIF
\ENDIF
\end{algorithmic}
\end{algorithm}

To verify that (\ref{eqn:likelihood}) is a good estimation, 
suppose we have processed $N$ blocks with depth $d$.
Notice that $T_d^+$ is updated when $d_f > d$, if we further consider $p$ and $X$ (Cf. line 5-9 in the pseudocode), 
writing $k_1=1$ and $k_2=\tau_2^{-1}$, we have
\begin{align*}
E(T_d^+(d_{f_L},d_{f_R})) &= Nk_1P(d_{f_L},d_{f_R}, d_f > d \given p > \tau_1)P(p > \tau_1)~+ \\
&Nk_2P(d_{f_L},d_{f_R}, d_f > d \given p \le \tau_1, X < \tau_2)P(p \le \tau_1 \given X < \tau_2)P(X < \tau_2) \\
&= NP(d_{f_L},d_{f_R}, d_f > d)
\end{align*}
as $X$ is indepedent and $k_2P(X<\tau_2)=1$. 
Then the first part of (\ref{eqn:likelihood}) holds assuming $N$ is sufficiently large. The second part of (\ref{eqn:likelihood}) 
can be proved in a similar fashion.

To see how $\tau_1, \tau_2$ affect the performance,
we treat the posterior probability $p$ as a random variable with density $g(p)$. 
Let $t_0$ and $b_0$ be the expected time and bitrate cost of fully encoding a block with depth $d$,
while the reduced expected time cost with early termination is $t$, and
the increased expected bitrate cost is $b$ if it should be 4-split.
Assuming independence between $p$, $t$, and $b$,
the expected time saving and bitrate cost increment (in percentage) of encoding $N$ blocks with depth $d$ comparing to the original encoder is
\begin{align} \label{eqn:delta}
\Delta T = \frac{N_t(t-t_0)}{Nt_0},~
\Delta B = \frac{N_b(b-b_0)}{Nb_0},
\end{align}
where $N_t$ is the expected number of blocks that are not fully encoded, 
and $N_b$ is the expected number of blocks that are not optimally partitioned:
\begin{align} \label{eqn:nt-nb}
N_t = N\int_0^{\tau_1}g(p)(1-\tau_2)\mathrm{d}p,~
N_b = N\int_0^{\tau_1}pg(p)(1-\tau_2)\mathrm{d}p < \tau_1N_t,
\end{align}
define $\Delta t=(t-t_0)/t_0 < 0$ and $\Delta b=(b-b_0)/b_0 > 0$, (\ref{eqn:delta})(\ref{eqn:nt-nb}) imply that
\begin{align} \label{eqn:deltat-deltab}
\Delta T = \Delta t(1-\tau_2)\int_0^{\tau_1}g(p)\mathrm{d}p,~
\frac{\Delta B}{\Delta b} < \tau_1 \frac{\Delta T}{\Delta t} = \tau_1 (1-\tau_2)\int_0^{\tau_1}g(p)\mathrm{d}p.
\end{align}
We see from (\ref{eqn:deltat-deltab}) that a larger $\tau_1$ and/or smaller $\tau_2$ gives better time saving,
and the bitrate penalty is bounded above by the time saving, with a multiplier of $\tau_1$. 
It also implies that a smaller $\tau_1$ results in less coding efficiency deterioration.

Finally, to see the effect of $\tau_2$ on (\ref{eqn:likelihood}), consider the case $p\le \tau_1$. 
The increment to $T_d^+$ (or $T_d^-$) is a Bernoulli random variable $\delta \sim \tau_2^{-1}\text{Bern}(\tau_2)$. 
We see that $E(\delta) = 1$ and $\mathrm{Var}(\delta) = \tau_2^{-1}-1$.
In conclusion, $\tau_1, \tau_2$ should be chosen carefully to achieve a good balance between
time saving, bitrate penalty, and statistical stabilization.

\begin{figure}[t]
\begin{center}
\begin{tabular}{cc}
\epsfig{width=2.5in,file=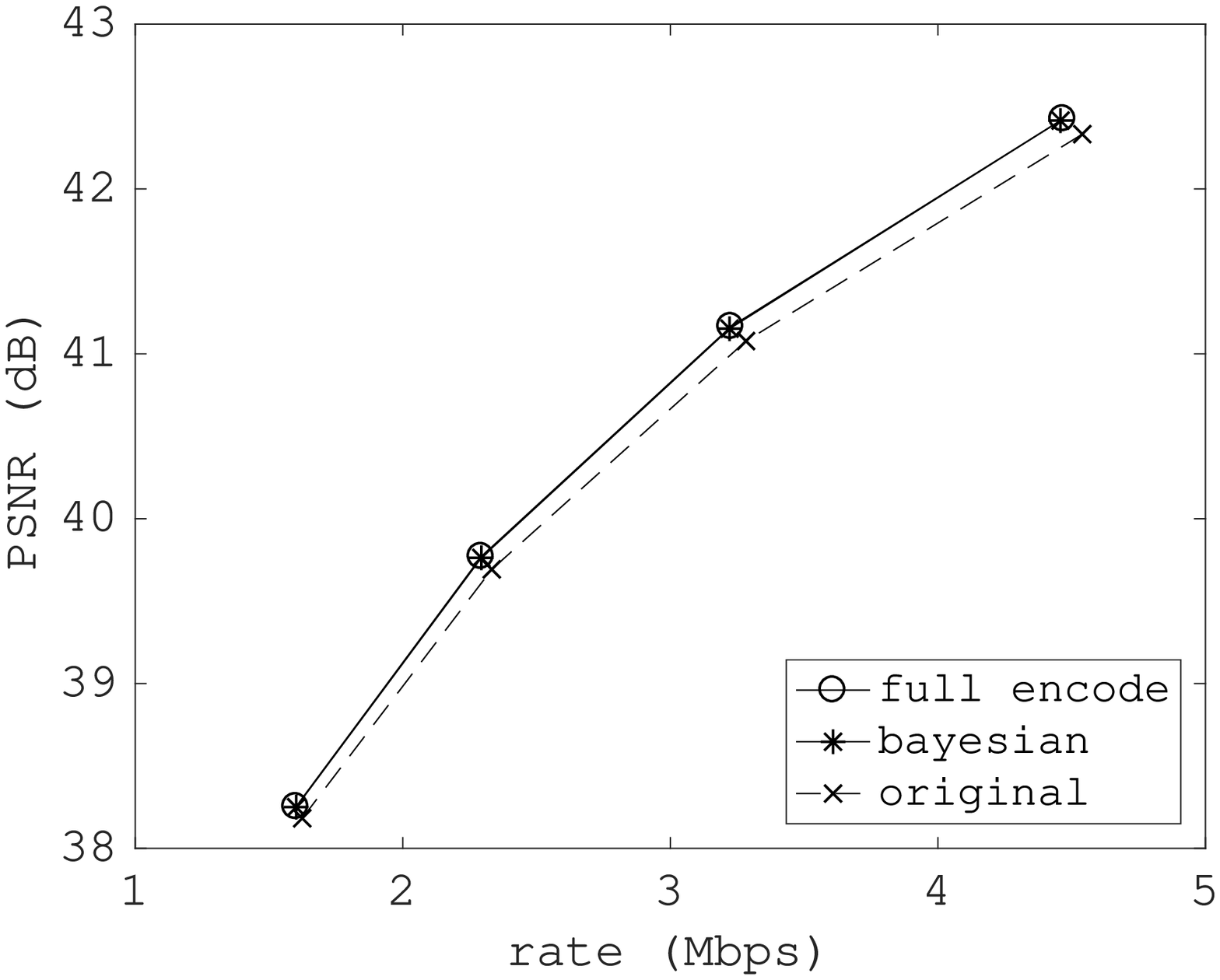} &
\epsfig{width=2.5in,file=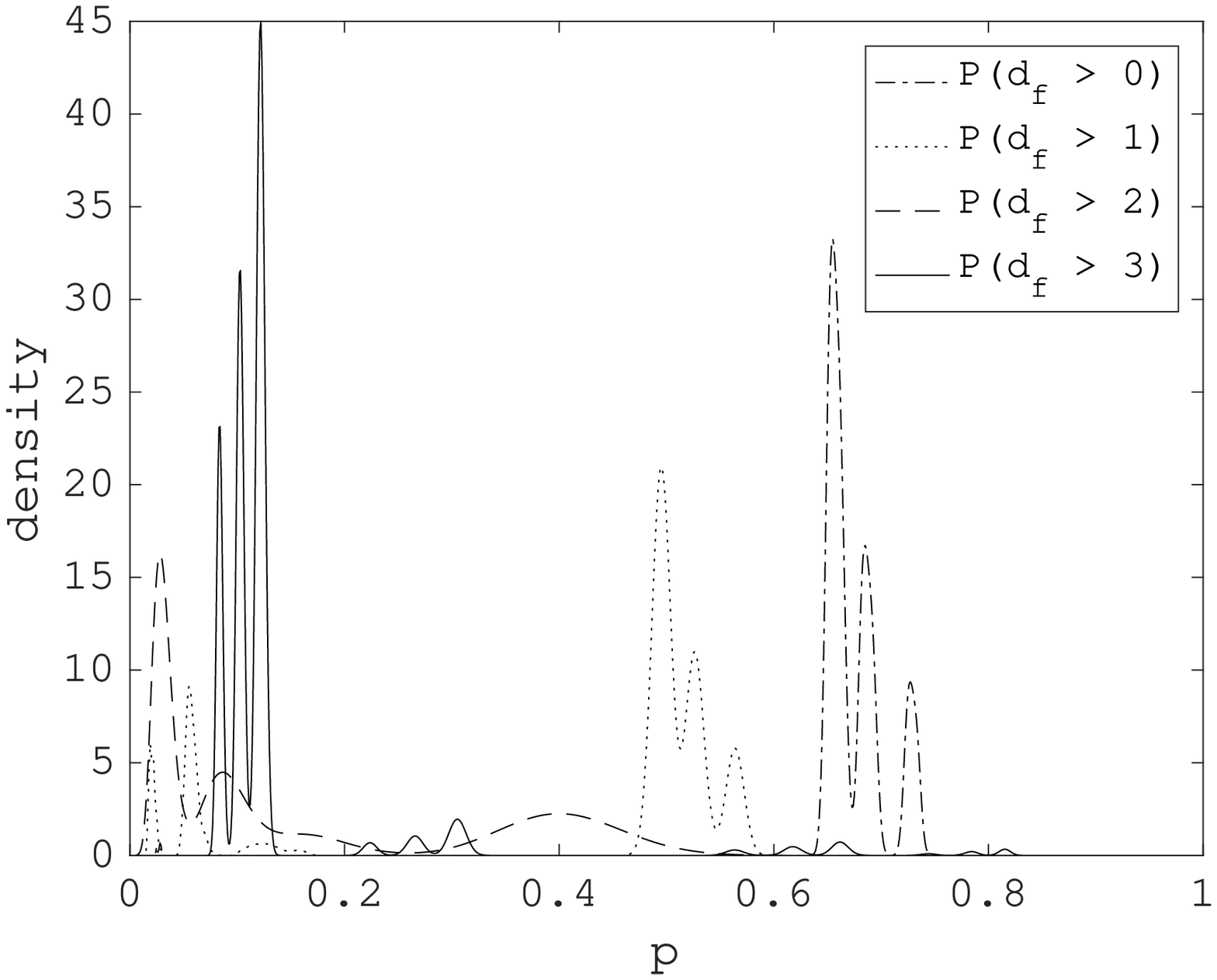} \\
{\small (a)} & {\small (b)}
\end{tabular}
\end{center}
\caption{\label{fig:experimental-results-figure}%
RD curves and the posterior distribution}
\end{figure}

\Section{Experimental Results}
Table~\ref{tab:bayesian-experiment} shows the encoding results of our proposed bayesian method,
using the same test setup as Table~\ref{tab:simple-rdo-pruning}. 
We set $\tau_2=0.05$ as the analysis in the previous section shows that smaller (but non-zero) $\tau_2$ is preferred for better time savings.
We varied the value of  $\tau_1$ to examine its impact on the performances.
Fig.~\ref{fig:experimental-results-figure}(a) compares the RD performances of the original pruning algorithm 
and our bayesian method regarding the sequence \textit{BasketballDrill} where $\tau_1$ is set to $0.2$.

\begin{table}[tp]
\begin{center}
\caption{\label{tab:bayesian-experiment}%
Experimental results of the bayesian method}
{
\renewcommand{\baselinestretch}{1}\footnotesize
    \centering
        \begin{tabular}{|l|c|c|c|c|c|c|}
        \hline
         & \multicolumn{3}{|c|}{BD-rate} &\multicolumn{3}{|c|}{$\Delta T$} \\
        \cline{2-7}
        Sequence &$\tau_1=0.1$ &$\tau_1=0.2$ &$\tau_1=0.4$ &$\tau_1=0.1$ &$\tau_1=0.2$ &$\tau_1=0.4$ \\
        \hline
        BasketballPass (412$\times$240)  & 0.13\% & 0.20\% & 0.74\% & -22.5\% & -24.7\% & -26.7\%\\
        BlowingBubbles (412$\times$240)  & 0.16\% & 0.24\% & 0.37\% & -24.9\% & -26.3\% & -29.1\%\\
        BQSquare (412$\times$240)        & 0.00\% & 0.13\% & 0.46\% & -18.0\% & -23.7\% & -25.7\%\\
        \hline
        BasketballDrill (832$\times$480) & 0.15\% & 0.22\% & 0.58\% & -30.5\% & -37.8\% & -38.4\%\\
        PartyScene (832$\times$480)      & 0.03\% & 0.20\% & 0.73\% & -20.2\% & -24.0\% & -28.8\%\\
        \hline
        FourPeople (1280$\times$720)     & 0.03\% & 0.06\% & 0.07\% & -18.6\% & -42.8\% & -48.2\%\\
        Johnny (1280$\times$720)         & 0.09\% & 0.10\% & 0.25\% & -18.8\% & -37.0\% & -50.6\%\\
        \hline
        Kimono (1920$\times$1080)        & 0.04\% & 0.07\% & 0.14\% & -10.4\% & -25.4\% & -41.5\%\\
        \hline
        \textbf{Average}                & \textbf{0.08\%} & \textbf{0.15\%} & \textbf{0.46\%} & \textbf{-20.5\%} & \textbf{-30.2\%} & \textbf{-36.1\%}\\
        \hline
        \end{tabular}
}
\end{center}
\end{table} 

We see from the results that our proposed method achieves significant time savings (up to $50.6\%$) 
without substantial degradation to coding efficiencies.
An interesting pattern is that,
sequences with higher spatial resolution can tolerate a larger $\tau_1$ in terms of BD-rate cost, 
in exchange of a significant margin in time savings,
while for sequences with lower spatial resolutions this no longer holds,
as a smaller $\tau_1$ is clearly more favorable.
This can be explained by the depths of prime blocks in high resolution sequences being relatively lower
(Cf. Fig.~\ref{fig:block-structure-statistics}(c)),
encouraging early terminations in the RDO procedure.
Nevertheless, setting a constant $\tau_1=0.4$ for all sequences yields an average time saving of $36.1\%$ 
(and even higher for high resolution sequences),
with an average BD-rate cost of $0.46\%$, which is negligible for most practical applications.

\begin{table}[tp]
\begin{center}
\caption{\label{tab:revised-vs-bayesian}%
Controlled study of two proposed methods}
{
\renewcommand{\baselinestretch}{1}\footnotesize
    \centering
        \begin{tabular}{|l|c|c|c|c|c|}
        \hline
         && \multicolumn{2}{|c|}{BD-rate} &\multicolumn{2}{|c|}{$\Delta T$} \\
        \cline{3-6}
        Sequence & $\tau_1$ & improved & bayesian & improved & bayesian \\
        \hline
        BasketballPass (412$\times$240)  & 0.1 & 0.15\% & 0.13\% & -18.9\% & -22.5\%\\
        BlowingBubbles (412$\times$240)  & 0.1 & 0.25\% & 0.16\% & -21.1\% & -24.9\%\\
        BQSquare (412$\times$240)        & 0.1 & 0.05\% & 0.00\% & -17.6\% & -18.0\%\\
        \hline
        BasketballDrill (832$\times$480) & 0.2 & 0.16\% & 0.22\% & -30.6\% & -37.8\%\\
        PartyScene (832$\times$480)      & 0.2 & 0.08\% & 0.20\% & -19.8\% & -24.0\%\\
        \hline
        FourPeople (1280$\times$720)     & 0.4 & 0.10\% & 0.07\% & -40.2\% & -48.2\%\\
        Johnny (1280$\times$720)         & 0.4 & 0.09\% & 0.25\% & -42.4\% & -50.6\%\\
        \hline
        Kimono (1920$\times$1080)        & 0.4 & 0.24\% & 0.14\% & -37.5\% & -41.5\%\\
        \hline
        \multicolumn{2}{|c|}{\textbf{Average}} & \textbf{0.14\%} & \textbf{0.14\%} & \textbf{-28.5\%} & \textbf{-33.4\%} \\
        \hline
        \end{tabular}
}
\end{center}
\end{table} 

The bayesian approach offers the flexibility to control the tradeoff between time savings and bitrate cost,
as shown in Table~\ref{tab:bayesian-experiment}.
Furthermore, Table~\ref{tab:revised-vs-bayesian} shows the advantage of the joint inference model over the improved pruning method, 
for controlling the BD-rate cost.
Larger $\tau_1$ values are chosen for higher resolution sequences,
as explained in the previous paragraph.
In many cases both BD-rate costs and time savings are improved, without having to trade one for another, 
and on average time saving can be achieved without BD-rate loss.

Finally, we briefly discuss the statistical behavior of the posterior, previously defined as $g(p)$.
Fig.~\ref{fig:experimental-results-figure}(b) shows an estimated $g(p)$ when encoding \textit{BasketballDrill} under $Q=32$.
The bayesian method essentially classifies blocks according to their computed posteriors.
According to (\ref{eqn:deltat-deltab}),
the portion of $g(p)$ where $p < \tau_1$ is eligible for RDO early termination.
This implies that a posterior admitting bimodal distribution is ideal for our purpose.
Many aspects can be incorporated into the posterior model for further optimization, 
including spatial resolution, quality levels, RD information, etc, which however, is beyond the scope of this paper.

\Section{Conclusions}
In this paper, we propose a novel bayesian block structure inference framework inspired by a modification to an existing HEVC-based 
multi-rate encoding algorithm. 
The proposed method is effective, flexible and especially suitable for codecs utilizing special reference frames like AV1.
Experimental results show that the proposed bayesian algorithm can achieve $36.1\%$ time savings in average 
and up to $50.6\%$, while keeping the bitrate cost below negligible level.

Future work includes optimizations to the posterior model,
dynamic adaptation of $\tau_1$,
adopting the bayesian approach to prediction mode inference, 
motion vector/intra mode inference, 
as well as multi-resolution encoding.

\Section{Acknowledgements}
This work was supported by the Natural Science Foundation of China
(Project Number 61521002).
\Section{References}
\bibliographystyle{IEEEtran}
\bibliography{refs}

\end{document}